\documentclass{article}

\usepackage{arxiv}

\usepackage[utf8]{inputenc} 
\usepackage[T1]{fontenc}    
\usepackage{url}            
\usepackage{booktabs}       
\usepackage{amsfonts}       
\usepackage{nicefrac}       
\usepackage{microtype}      
\usepackage{lipsum}
\usepackage{graphicx}
\graphicspath{ {./images/} }
\usepackage{graphicx}
\usepackage{amsmath}
\usepackage{amssymb}
\usepackage{natbib} 

\usepackage{newtxtext}
\usepackage{newtxmath}
\usepackage{bm}
\usepackage{marvosym}

\usepackage[
    colorlinks=true, 
    linkcolor=black, 
    citecolor=black  
]{hyperref}

\title{Vortical similarities across laminar and turbulent extreme gust encounters}

\author{
    Hiroto Odaka\thanks{Email for correspondence: hodaka@g.ucla.edu}, \  Barbara Lopez-Doriga, \ and Kunihiko Taira \\ \\
    Department of Mechanical and Aerospace Engineering, University of California, Los Angeles, CA 90095, USA
}

\begin{document}
\maketitle
\begin{abstract}
This study uncovers a striking similarity between massively separated laminar and turbulent flows that develop over a square wing during extreme vortex gust encounters.
The evolving large-scale, vortical core structures responsible for significant transient lift variations exhibit remarkable similarity across $Re=600$ and 10,000. 
The formation of these structures is attributed to a substantial gust-induced vorticity flux produced at the wing surface, resulting in shared large-scale topological features between the low- and high-Reynolds-number flows.
Although fine-scale vortical structures quickly emerge in the $Re=$ 10,000 case, the large-scale structures identified by scale decomposition of the turbulent flow resemble those observed at $Re=600$.
These findings suggest that large-scale vortical features present in laminar extreme aerodynamic flows provide key insights into their higher Reynolds number counterparts, potentially reducing the complexity of flow modeling and control for extreme aerodynamics.
\end{abstract}


\section{Introduction}
\label{sec:intro}

Modern small-scale aircraft, such as drones, are increasingly utilized in a range of operations, including logistics, emergency response, infrastructure inspection, and national security.
However, aerodynamically challenging flight environments have been reluctantly avoided~\citep{anya_review,gianfelice2022real,mohamed2023gusts}.
In particular, airspace with high occurrences of gusts, such as urban canyons and mountainous areas, is hazardous because the gust disturbances can be relatively strong due to the small size and low velocity of the aircraft \citep{mueller2003aerodynamics,anya_review}.
Such strong gust encounters lead to unsteady, complex dynamics with strong nonlinearities in the flowfield around the vehicles and exert significantly large transient forces on their bodies.

To deepen our understanding of aerodynamics in such severely gusty environments, there have been recent studies in \textit{extreme aerodynamics}, where the gust ratio~$G$, the ratio between gust velocity and free-stream velocity, is larger than~1~\citep{anya_review,taira2025extreme}.
Recent studies have analyzed and modeled two-dimensional extreme vortex-gust encounters by an airfoil at~$Re=100$ \citep{fukami2023grasping,fukami2024data}.  In these studies, a machine-learned model was used to implement lift attenuation control in low-dimensional latent space during the encounters. 
Building on this latent space model, \citet{mousavi2025low} performed flow estimation with uncertainty quantification using sparse sensor measurements.
Furthermore, \citet{fukami2025extreme} presented machine-learning-based data compression of extreme gust encounters by a spanwise periodic wing at a Reynolds number of~$5000$.
To further investigate wingtip effects on the extreme aerodynamic events, \citet{odaka2025extreme} conducted direct numerical simulations~(DNS) of extreme vortex-gust encounters by a square wing at~$Re=600$ and analyzed the complex three-dimensional vortex dynamics and the modified lift loads.

Most of these studies in extreme aerodynamics with massive flow separation have been performed at low Reynolds numbers \citep{jones2020gust,taira2025extreme}.  
A critical question that follows is: 
How can we apply insights from massively separated flows in the laminar regime of extreme aerodynamics to their turbulent regime counterpart?
The current study answers this question by performing numerical simulations of extreme vortex-gust-square-wing interaction using DNS at~$Re=600$ and large eddy simulation~(LES) at~$Re=$ 10,000. 
We in particular focus on the vortex dynamics and the pressure evolution around the wing in the extreme flows.
We further identify vortical structures that play aerodynamically important roles in the unsteady flowfield using the force element method \citep{chang1992potential}, showing that the prominent structures are shared across the low and high Reynolds numbers.
Moreover, scale decomposition analysis \citep{motoori2019generation,fujino2023hierarchy} is performed on the $Re=$10,000 flow to extract large-scale, dominant flow structures, where a quantitative similarity to those observed at $Re=600$ is uncovered.
This work shows that insights from laminar extreme aerodynamics can be extended to turbulent cases, suggesting that we can potentially simplify flow analysis, modeling, and control for turbulent extreme aerodynamics.

The remainder of the paper is organized as follows. 
The problem setup for the extreme vortex-gust-square-wing interaction for a gust ratio of~$\pm2$ is provided in \S\ref{sec:problemSetup}. 
In \S\ref{sec:result}, we present vortex dynamics and identify vortical structures responsible for large transient lift change for both laminar and turbulent conditions at $Re=600$ and 10,000, respectively.
This study reveals striking similarity of large-scale flow structures between the two Reynolds numbers.
In \S\ref{sec:conclusion}, a summary of our findings is offered.

\section{Problem setup}
\label{sec:problemSetup}

\begin{figure}
\begin{center}
   \includegraphics[width=1\textwidth, scale=1.0]{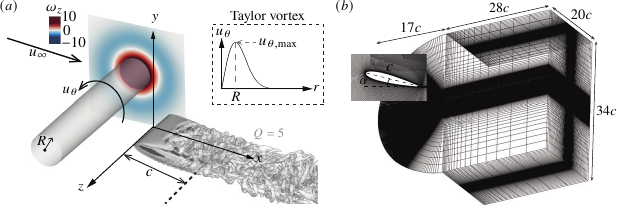}
   \caption{
   (a)~A square NACA0015 wing encountering a gust vortex.
   Q-criterion isosurface~$Q=5$ is shown.
   (b)~Computational domain and spatial discretization.
   }
\label{fig:fig1}
\end{center}
\end{figure}

We consider a vortex gust, modeled as a spanwise-oriented Taylor vortex \citep{Taylor_vortex}, impacting a full-span square wing (aspect ratio~$AR=1$), as shown in figure \ref{fig:fig1}$(a)$.
The azimuthal velocity of this vortex is
\begin{equation}
    u_{\theta}=u_{\theta, \text{max}} \frac{r}{R} \exp \left[ \frac{1}{2} \left( 1- \frac{r^2}{R^2} \right)  \right],
\end{equation}
where $u_{\theta, \text{max}}$ is the maximum velocity at radius $r=R$.
The profiles of the tangential velocity~$u_{\theta}$ and the spanwise vorticity~$\omega_z$ are also presented in figure \ref{fig:fig1}$(a)$.
The strength of this vortex is characterized by the gust ratio,
\begin{equation}
    G \equiv u_{\theta,\text{max}}/u_{\infty},
\end{equation}
where $u_{\infty}$ is the free-stream velocity.
We choose two gust ratios~$G= \{2,-2 \}$ and set the radius to $R/c=0.25$.
The square wing has a straight-cut tip with a chord length~$c$ and an angle of attack~$\alpha=14^\circ$.
We examine the flows at chord-based Reynolds numbers $Re=u_{\infty}c/\nu=600$ and 10,000, where~$\nu$ is the kinematic viscosity.
We select these values of $Re$ to characterize the extreme aerodynamic flows in laminar and turbulent regimes.

We perform DNS for~$Re=600$ and LES for~$Re=$ 10,000 with a compressible flow solver CharLES~\citep{charles2} -- a finite-volume solver with second-order accuracy in space and third-order accuracy in time.  For LES, the Vreman subgrid-scale model~\citep{vreman2004eddy} is used.
The Mach number~$u_\infty/a_\infty=0.1$ is set to minimize compressible effects.
The computational domain is presented in figure~\ref{fig:fig1}$(b)$.
We define $x$, $y$, and $z$ in the streamwise, transverse, and spanwise directions, respectively, with the leading edge of the wing root placed at the origin.
We prescribe adiabatic wall boundary condition at the wing surface, a Dirichlet boundary condition of $u_\infty/a_\infty=0.1$ at the inlet and farfield boundaries, and a sponge zone next to the outlet boundary.
At the spanwise boundaries along~$z/c=\pm10$, we apply a symmetry boundary condition to maintain the form of the gust vortex column far from the wing.

The grid and domain setups have been carefully verified.
The current mesh has $31$ million control volumes with $n_\text{airfoil}=480$, $n_z=90$, and $y^+_0=0.24$, where $n_\text{airfoil}$ and $n_z$ are the grid points along the wing tangent~(around the upper and lower surfaces) and the wingspan, and $y^+_0$ is the first grid off the wall surface in viscous wall units at~$Re=$ 10,000.
The refined mesh used for verification has $62$ million cells with $n_\text{airfoil}=760$, $n_z=144$, and $y^+_0=0.15$.  The time step is chosen to ensure that the local Courant number is smaller than~1 over the entire computational domain.
The current simulation setup has been carefully validated with~\citet{fukami2025extreme}.

The gust vortex is introduced at $(x_0/c,y_0/c)=(-3,0)$ in the undisturbed flow at a reference time of~$\tau=u_\infty t/c=-3$ and convects with the freestream.
Time of $\tau=0$ corresponds to the moment at which the center of the gust vortex would reach the leading edge of the wing if there were no wing.

\section{Results and discussion}
\label{sec:result}

\begin{figure}
\begin{center}
   \includegraphics[width=1\textwidth, scale=1.0]{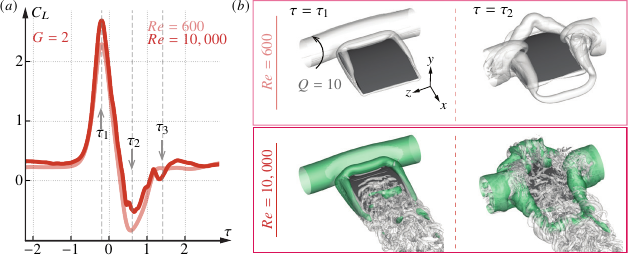}
   \caption{
   (a)~Lift history for the $G=2$ case at $Re=600$~(light red) and~10,000~(dark red).
   (b)~Top-port view of the Q-criterion isosurface~$Q=10$, colored in gray, at $\tau=\tau_1$ and $\tau_2$ noted in~(a).
   The Q-criterion isosurface~$Q=10$ of the large-scale structures extracted by the scale decomposition analysis with~$\sigma/c=0.05$ in the $Re=$ 10,000 case is superposed in green.
   }
\label{fig:fig2}
\end{center}
\end{figure}

We begin by presenting the lift fluctuation for the~$G=2$ case in the laminar and turbulent regimes.  In figure~\ref{fig:fig2}$(a)$, the light and dark red lines represent the lift histories for $Re=600$ and 10,000, respectively.
The lift coefficient is defined as $C_L=F_L/(\frac{1}{2} \rho u_\infty^2 bc)$, where $F_L$ is the lift force on the wing, $\rho$ is the fluid density, and $b$ is the span length.
Before the gust vortex is introduced in the flowfield, the flow is steady at $Re=600$ with $C_{L,\text{base}}=0.22$, while it is unsteady at $Re=$ 10,000 with a time-averaged lift coefficient of $\bar{C}_{L,\text{base}}=0.32$.
As the gust vortex approaches the wing, the wing experiences a substantial lift increase, which peaks at $\tau=\tau_1$ for both $Re=600$ and 10,000.
After the positive peak, the lift drops and reaches its negative peak at $\tau=\tau_2$.
Subsequent to these lift fluctuations, the lift recovers to the baseline values.

The time evolution of the vortex cores is presented in figure \ref{fig:fig2}$(b)$, where we visualize the Q criterion isosurface at $Re=600$ and 10,000 in gray.
For the $Re=$ 10,000 case, the Q criterion isosurface of large-scale structures extracted through scale decomposition \citep{motoori2019generation,fujino2023hierarchy} is also visualized in green, where we decompose a velocity component $u_i$ into the large-scale $\widetilde{u}_{i_L}$ and the small-scale~$\widetilde{u}_{i_S}$ by applying a Gaussian filter $K$ in three spatial directions such that
\begin{align}
        &\widetilde{u}_{i_L}(x,y,z; \sigma) = \sum_{x_p}\sum_{y_p}\sum_{z_p} u_i(x_p,y_p,z_p) K(x_p,y_p,z_p;x,y,z),\\
        &\widetilde{u}_{i_S}(x,y,z; \sigma) = u_i - \widetilde{u}_{i_L}(x,y,z; \sigma).
\end{align}
Here, $x_p$, $y_p$, and $z_p$ are the coordinates of grid points. 
The Gaussian kernel is
\begin{equation}
        K(x_p,y_p,z_p;x,y,z)= C(x,y,z) \exp{  \left[-\frac{(x_p-x)^2+(y_p-y)^2+(z_p-z)^2}{2\sigma^2}\right]} \Delta x \Delta y \Delta z
\end{equation}
and the coefficient $C(x,y,z)$ is selected to ensure the sum of the kernel is unity; \begin{equation}
\sum_{x_p}\sum_{y_p}\sum_{z_p} K(x_p,y_p,z_p;x,y,z)=1.
\end{equation}
Parameter~$\sigma$ is the effective size of the Gaussian filter, where information approximately smaller than $\sigma$ is smoothed out in the filtered velocity~$\widetilde{u_i}_{|L}$~\citep{motoori2019generation}. 
Here, $\sigma/c=0.05$ is chosen to uncover the large-scale vortical structures that are primarily responsible for the lift peaks, shown in figure~\ref{fig:fig2}$(b)$.  
The influence of $\sigma/c$ is examined later.

As shown in figure \ref{fig:fig2}$(b)$, the large-scale flow structures between the laminar and turbulent flows are strikingly similar.
At the first lift peak $\tau=\tau_1$, we observe the development of a large leading-edge vortex~(LEV) for both $Re=600$ and 10,000.
This results from a substantial amount of gust-induced wall-normal vorticity production at the leading edge \citep{taira2025extreme}.
Notably, these accentuated vorticity production levels at the leading edge are directly related to the enhanced surface pressure gradients caused by the incoming gust vortex \citep{lopez2025effect}.
During the gust-leading-edge interaction, which is up until the impingement, viscous effects are not prominent \citep{peng2015vortex}, and similar large-scale topological features are observed across the two Reynolds numbers.
Even at~$Re=$ 10,000, the locally accelerated flow near the leading edge experiences stabilizing effects \citep{linot2024laminar}, producing large-scale, laminar flow structures.

Furthermore, the tip vortices~(TiVs) grow in size near the leading edge for both Reynolds-number flows during the encounter.
This is because the impacting gust vortex induces a large vorticity flux at the tip near the leading edge while increasing the pressure difference between the two sides of the wing~\citep{odaka2025extreme}.  
Similar to the LEV, the large-scale features of the TiVs are shared across the Reynolds numbers since the formation of the vortices is pressure-driven.
Around the negative lift peak~$\tau=\tau_2$, we observe the LEV transforms into an arch vortex and a large trailing-edge vortex~(TEV) shed for both $Re=600$ and 10,000.
Despite the presence of finer-scale vortical structures at $Re=$ 10,000, the large-scale structures are remarkably similar to those observed at~$Re=600$.

\begin{figure}
\begin{center}
   \includegraphics[width=1\textwidth, scale=1.0]{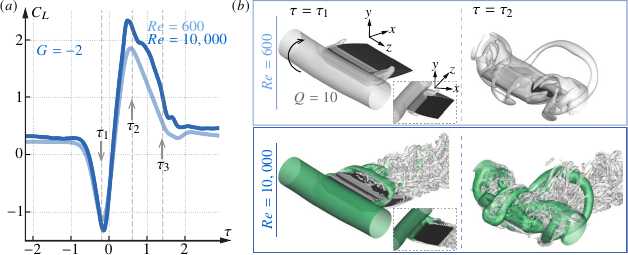}
   \caption{
   Same plot as figure~\ref{fig:fig2} for the negative gust vortex case with~$G=-2$.
   }
\label{fig:fig3}
\end{center}
\end{figure}

Next, let us examine the negative gust vortex case with~$G=-2$.  During the encounter, the wing experiences a sharp lift decrease with its negative peak at~$\tau=\tau_1$ followed by a rapid increase with its positive peak around $\tau=\tau_2$, as shown in  figure~\ref{fig:fig3}$(a)$.
Lift gradually recovers to its baseline value afterward.  These trends are shared across the laminar and turbulent flow cases.

In figure~\ref{fig:fig3}$(b)$, we visualize the Q criterion isosurface for the $G=-2$ cases at $Re=600$ and 10,000.
At~$\tau=\tau_1$, an LEV develops below the wing due to a large vorticity flux at the leading edge induced by the approaching negative gust vortex.
Furthermore, the impacting gust vortex increases the pressure on the upper surface while decreasing it below the wing, inverting the suction and pressure sides of the wing.
This causes the TiVs to weaken and even reverse their orientation~\citep{odaka2025extreme}.
As discussed in the positive gust vortex case, the gust-induced, large wall-normal vorticity production and flow acceleration near the leading edge lead to these similar core vortical structures across the Reynolds numbers.

Over time, the LEV below the wing forms an arch vortex, while the gust vortex convects above the wing.
At the positive lift peak around~$\tau=\tau_2$, the legs of the arch vortex are pushed toward the wing root, transforming into a hairpin vortex~\citep{odaka2025extreme}.
Although there is an increased number of finer structures around the wing in the turbulent flow, the hairpin vortex is clearly observed in the large-scale structures at~$Re=$ 10,000.
Again, we observe strong similarities in large-scale vortical structures across the two Reynolds numbers throughout the encounter.

\begin{figure}
\begin{center}
   \includegraphics[width=1\textwidth, scale=0.9]{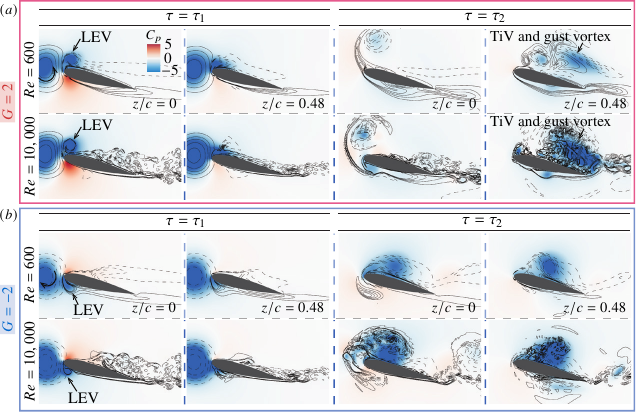}
   \caption{
   Spanwise slices of $C_p$ (color contours) with $\omega_z$ (line contours) along the root~$z/c=0$ and near the tip $z/c=0.48$ at $\tau=\tau_1$ and $\tau_2$ for (a) $G=2$ and (b) $G=-2$.
   Dashed line contours indicate negative $\omega_z$.
   }
\label{fig:fig4}
\end{center}
\end{figure}

Furthermore, we confirm strong similarity in the pressure fields at $Re=600$ and 10,000 influenced by their respective large-scale vortical structures.
In figure \ref{fig:fig4}, we show the spanwise slices of the pressure coefficient $C_p \equiv {(p-p_{\infty})}/{(\frac{1}{2} \rho u_\infty^2) }$
and the spanwise vorticity $\omega_z$ along the wing root $z/c=0$ and near the tip $z/c=0.48$, where $p_{\infty}$ is the farfield pressure. 
For the $G=2$ case (figure \ref{fig:fig4}$(a)$), the LEV carries a low-pressure core around the leading edge at~$\tau=\tau_1$ at both $Re$.
Moreover, the stronger spanwise vorticity of the gust vortex at~$\tau=\tau_1$ for the $Re=$ 10,000 case is observed in the denser contour lines of $\omega_z$, because there is lower viscous diffusion at the higher Reynolds number, enabling the gust vortex to retain higher intensity in the spanwise vorticity.
This leads to a higher first lift peak at~$\tau=\tau_1$ for~the $Re=$ 10,000 case than that for~the $Re=600$ case, as shown in figure~\ref{fig:fig2}$(a)$.

At~$\tau=\tau_2$ for the $G=2$ case, the lowest-pressure regions are above the wing near the tips for both laminar and turbulent flows, as shown in figure \ref{fig:fig4}$(a)$.
These regions correspond to the TiVs and the gust vortex near the wingtips, as seen in the visualized Q criterion at $\tau=\tau_2$ in figure~\ref{fig:fig2}$(b)$.
The $Re=$ 10,000 case exhibits lower-pressure regions above the wing near the tips, resulting in a smaller negative lift peak compared to the $Re=600$ case, as seen in figure \ref{fig:fig2}$(a)$.
At $Re=$ 10,000, fine-scale structures are present over a broad region above the wing, but are absent below it at $\tau=\tau_2$, as seen in the contour lines of $\omega_z$ in figure \ref{fig:fig4}$(a)$.
This is due to the flow below the wing having been accelerated by the lower portion of the gust vortex.
As discussed previously, accelerated flows have the tendency to be stabilized with coherent, laminar features, whereas decelerated flows tend to trigger instabilities with flow structures broken into finer structures \citep{linot2024laminar}.

For the negative gust vortex case with $G=-2$ shown in figure \ref{fig:fig4}$(b)$, the LEV below the wing at $\tau=\tau_1$ and the gust vortex convecting above the wing at $\tau=\tau_2$ create low-pressure cores at both Reynolds numbers.
At $\tau=\tau_2$, the intensity of the gust vortex convecting above the wing is larger in the turbulent case than in the laminar case, due to lower vortex diffusion at the higher Reynolds number.
This leads to lower-pressure regions above the wing for the $Re=$ 10,000 flow compared to the $Re=600$ flow at $\tau=\tau_2$ (figure \ref{fig:fig4}$(b)$), contributing to the larger second lift peak seen in figure \ref{fig:fig3}$(a)$.
These findings suggest that the effects of extreme gusts are pressure-dominated, driven by large-scale vortices, whose primary dynamics are similar across laminar and turbulent regimes.

Next, let us extract the vortical structures responsible for the large lift change during the encounters.
To identify vortical structures responsible for the lift, we use the force element analysis~\citep{chang1992potential}.
An auxiliary potential $\phi_y$ is defined for a boundary condition of~$-\boldsymbol{n}\cdot\nabla \phi_y=\boldsymbol{n}\cdot\boldsymbol{e}_y$ on the wing surface.
Here, $\boldsymbol{e_y}$ is the unit vector in the $y$ direction.
Integrating the inner product of the Navier-Stokes equations with the gradient of the auxiliary potential over the fluid domain, the lift force can be expressed as
\begin{equation}
    F_L=\int_V\boldsymbol{\omega}\times\boldsymbol{u}\cdot\nabla\phi_y dV + \frac{1}{Re}\int_S \boldsymbol{\omega}\times\boldsymbol{n}\cdot \left( \nabla\phi_y+\boldsymbol{e}_y \right)dS.
\end{equation}
The first and second integrands represent volume and surface lift elements, respectively.
The unsteady lift force is primarily attributed to the volume lift elements for the present flows at $Re = 600$ and 10,000 \citep{ribeiro2023laminar,odaka2025extreme}.
As such, the volume lift element is hereafter referred to as lift element $L_e$.

\begin{figure}
\begin{center}
   \includegraphics[width=1\textwidth, scale=0.9]{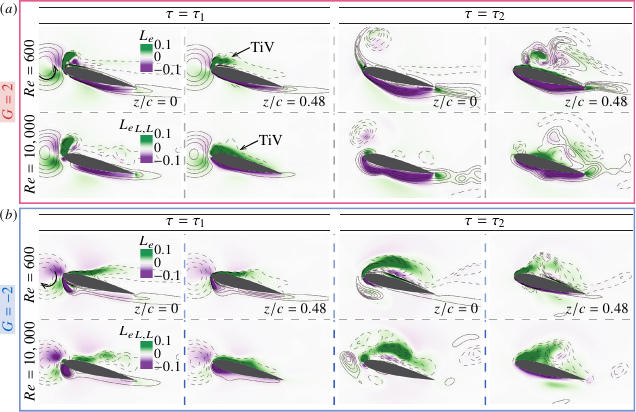}
   \caption{
   Spanwise slices of~$L_e$~(color contours) with~$\omega_z$~(line contours) at $Re=600$ and ${L_e}_{L,L}$ with~$\widetilde{{\omega}}_{z_L}$ extracted with $\sigma/c=0.05$ in the $Re=$ 10,000 flow along the root~$z/c=0$ and near the tip~$z/c=0.48$ at~$\tau=\tau_1$ and~$\tau_2$ for (a)~$G=2$ and (b)~$G=-2$.
   Dashed contour lines indicate negative~$\omega_z$ and $\widetilde{{\omega}}_{z_L}$.
   }
\label{fig:fig5}
\end{center}
\end{figure}

We can further decompose the lift element field for the~$Re=$ 10,000 case such that
\begin{equation}
\begin{split}
    \boldsymbol{\omega}&\times\boldsymbol{u}\cdot\nabla\phi_y=
\left(\widetilde{\boldsymbol{\omega}}_{L}+\widetilde{\boldsymbol{\omega}}_{S}\right) \times\left(\widetilde{\boldsymbol{u}}_{L}+\widetilde{\boldsymbol{u}}_{S}\right)\cdot\nabla\phi_y \\
&=
\underbrace{\widetilde{\boldsymbol{\omega}}_{L}\times\widetilde{\boldsymbol{u}}_{L}\cdot\nabla\phi_y}_{{L}_{e_{L,L}}}
+
\underbrace{(\widetilde{\boldsymbol{\omega}}_{S}\times\widetilde{\boldsymbol{u}}_{L})\cdot\nabla\phi_y}_{{L}_{e_{S,L}}}
+
\underbrace{(\widetilde{\boldsymbol{\omega}}_{L}\times\widetilde{\boldsymbol{u}}_{S})\cdot\nabla\phi_y}_{{L}_{e_{L,S}}}
+
\underbrace{\widetilde{\boldsymbol{\omega}}_{S}\times\widetilde{\boldsymbol{u}}_{S}\cdot\nabla\phi_y}_{{L}_{e_{S,S}}},
\end{split}
\label{eq:LE_decompose}
\end{equation}
where~$\widetilde{\boldsymbol{\omega}}_{L}=\nabla\times\widetilde{\boldsymbol{u}}_{L}$ and~$\widetilde{\boldsymbol{\omega}}_{S}=\nabla\times\widetilde{\boldsymbol{u}}_{S}$.
The terms of ${L_e}_{L,L}$, ${L_e}_{S,L}$, ${L_e}_{L,S}$, and ${L_e}_{S,S}$ represent the lift contribution from interactions between i) large-scale vorticity and large-scale velocity, ii) small-scale vorticity and large-scale velocity, iii) large-scale vorticity and small-scale velocity, and iv) small-scale vorticity and small-scale velocity, respectively.
Here, we expect dominance of ${L_e}_{L,L}$ over ${L_e}_{S,S}$, since large-scale vortices are the primary structures.
In addition, we expect the contribution of ${L_e}_{S,L}$ (small-scale vorticity and large-scale velocity) to be larger than that of ${L_e}_{L,S}$ (large-scale vorticity and small-scale velocity) because the vorticity distribution is more spatially compact than the velocity profile.

\begin{figure}
\begin{center}
   \includegraphics[width=1\textwidth, scale=0.9]{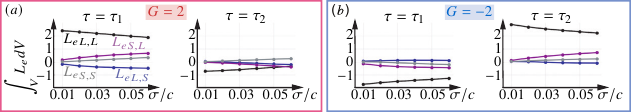}
   \caption{
   Volume integral of ${L_e}_{L,L}$, ${L_e}_{S,L}$, ${L_e}_{L,S}$, and ${L_e}_{S,S}$ at $\tau=\tau_1$ and $\tau_2$ over $\sigma/c$ for (a) $G=2$ and (b)~$G=-2$ at $Re=$ 10,000.
   }
\label{fig:fig6}
\end{center}
\end{figure}

In figure~\ref{fig:fig5}, the spanwise slices of~$L_e$ and~$\omega_z$ at $Re=600$ and ${L_e}_{L,L}$ and~$\widetilde{{\omega}}_{z_L}$ extracted with $\sigma/c=0.05$ at $Re=$ 10,000 along the wing root~$z/c=0$ and near the tip~$z/c=0.48$ are shown for $G=2$ and $-2$.
For the positive gust vortex case shown in figure~\ref{fig:fig5}$(a)$, the LEV and TiVs are responsible for lift generation at~$\tau=\tau_1$ in both laminar and turbulent flows.
Similar to the discussion in our previous work~\citep{odaka2025extreme}, the enhanced TiVs at~$\tau=\tau_1$ have large low-pressure cores above the wing, contributing to lift near the tips.
At the negative lift peak around~$\tau=\tau_2$, the bottom-surface boundary layer strongly contributes to negative lift.

We further examine the dominant structures in terms of lift production for the negative gust vortex case shown in figure~\ref{fig:fig5}$(b)$.
The LEV below the wing contributes to the negative lift peak at~$\tau=\tau_1$.
On the other hand, the positive lift peak around~$\tau=\tau_2$ is attributed to the vortical structure generated from the leading edge and the gust vortex convecting above the wing.
These coherent large-scale structures responsible for the lift peaks are similar between the low- and high-Reynolds number flows.

Let us also note that the extracted large-scale vortical structures at~$Re=$ 10,000 are primarily responsible for the significant lift peaks during the gust encounters.
In figure~\ref{fig:fig6}, we present the volume integral of each term in equation~\ref{eq:LE_decompose} at~$\tau=\tau_1$ and $\tau_2$ for~$G=2$ and $-2$ with varying~$\sigma/c$.
Here, the lift contribution from ${L_e}_{L,L}$ accounts for the majority of the lift peak values at~$\tau=\tau_1$ and~$\tau_2$ for both~$G=2$ and~$-2$.
Note that, for the minimum lift peaks at~$\tau=\tau_2$ for $G=2$ and at~$\tau=\tau_1$ for $G=-2$, the total lift values are negative, to which ${L_e}_{L,L}$ contributes the most.
This means that the extracted large-scale structures are the main contributors to the significant lift changes during the extreme gust encounters in the turbulent flow.

As $\sigma$ increases, a wider range of scales is incorporated as ``small-scale'' structures, while the coverage of ``large-scale'' structures decreases.
Consequently, as $\sigma$ increases, the contribution of~${L_e}_{L,L}$ decreases, while that of~${L_e}_{S,S}$ increases, as seen for both gust ratios at~$\tau=\tau_1$ and~$\tau_2$ in figure~\ref{fig:fig6}.
Furthermore, the contribution of~${L_e}_{S,L}$ rises as~$\sigma$ increases. 
This term holds components based on the spatially compact vorticity field with spatially non-compact velocity field.
With increased value of $\widetilde{\boldsymbol{\omega}}_{S}$ for a larger~$\sigma$, the contribution of~${L_e}_{S,L}$ increases slowly, even with more components being smoothed out for~$\widetilde{\boldsymbol{u}}_{L}$.
For instance, at~$\tau=\tau_2$ for the $G=2$ case~(figure~\ref{fig:fig6}$(a)$), while ${L_e}_{L,L}$ remains the largest contributor among the four terms for $\sigma/c \lesssim 0.05$, ${L_e}_{S,L}$ becomes the most dominant past $\sigma/c \approx 0.06$.
On the contrary, the contribution of~${L_e}_{L,S}$ does not grow as much with~$\sigma$.
The larger $\sigma$ is, the less local motion of the fluid is contained in~$\widetilde{\boldsymbol{u}}_{L}$, causing $\widetilde{\boldsymbol{\omega}}_{L}$ to decrease significantly.  Physically speaking, there cannot be a non-compact vorticity field producing a compact velocity field.
As a result, the lift contribution of~${L_e}_{L,S}$ remains low.

Based on the observation of ${L_e}_{L,L}$ being the most responsible for the lift peaks at $\sigma/c \lesssim 0.05$ with both gust ratios, we have visualized the flowfield evaluated on~$\widetilde{\boldsymbol{u}}_{L}$ with $\sigma/c=0.05$ in figures~\ref{fig:fig2}, \ref{fig:fig3}, and~\ref{fig:fig5}.
These findings show that the large-scale vortical structures in extreme gust encounters in turbulent regimes of at least~$\sim\mathcal{O}(10^4)$ are not only similar to those observed in the laminar regimes, but also primarily responsible for the substantial transient lift changes, suggesting that studies in extreme aerodynamics at low Reynolds numbers can provide critical insights into those at higher Reynolds numbers.

\begin{figure}
\begin{center}
   \includegraphics[width=1\textwidth, scale=1]{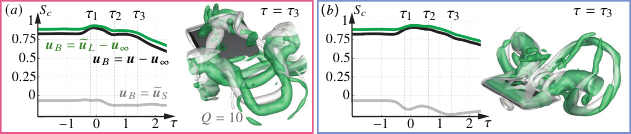}
   \caption{
   Time history of $S_c$ for (a) $G=2$ and (b) $G=-2$.
   The Q criterion isosurface of the $Re=600$ case~(gray) and extracted large-scale structures of the $Re=$ 10,000 case with $\sigma/c=0.05$ (green) at $\tau=\tau_3$ are visualized on the right of each subplot.
   }
\label{fig:fig7}
\end{center}
\end{figure}

Next, to mathematically quantify the similarity between these large-scale structures, we assess the similarity in the velocity field between the flows at $Re=600$ and 10,000.  
Here, we use the cosine similarity in $L^2$ between velocity fields~$\boldsymbol{u}_A$ and $\boldsymbol{u}_B$ defined as
\begin{equation}\label{eq:IP}
    S_c(\boldsymbol{u}_A,\boldsymbol{u}_B) = \frac{\int_V\boldsymbol{u}_A \cdot \boldsymbol{u}_B dV}{({\int_V||\boldsymbol{u}_A||^2dV})^{1/2}
    ({\int_V||\boldsymbol{u}_B||^2dV})^{1/2}}.
\end{equation}
In this study, $V=[0,3]\times[-1,1]\times[-0.8,0.8]$ and we set  $\boldsymbol{u}_A=\boldsymbol{u}-\boldsymbol{u_\infty}$ in the $Re=600$ flow.
We present the time history of $S_c(\boldsymbol{u}_A,\boldsymbol{u}_B)$ in figure \ref{fig:fig7} with $\boldsymbol{u}_B={\boldsymbol{u}}-\boldsymbol{u_\infty}$, 
$\boldsymbol{u}_B=\widetilde{\boldsymbol{u}}_{L}-\boldsymbol{u_\infty}$, and $\boldsymbol{u}_B=\widetilde{\boldsymbol{u}}_{S}$ for the $Re=$ 10,000 flow with  $\sigma/c=0.05$ when filtering is considered.

High level of quantitative similarity $S_c$ of the $Re=$ 10,000 case to the $Re=600$ case is presented for both $G=2$ and $-2$ in figure \ref{fig:fig7}.
As shown, the similarity is attributed to the large-scale structures in the $Re=$ 10,000 flow, and not the small-scale structures.
$S_c$ with $\boldsymbol{u}_B={\boldsymbol{u}}-\boldsymbol{u_\infty}$ or $\boldsymbol{u}_B=\widetilde{\boldsymbol{u}}_{L}-\boldsymbol{u_\infty}$
rises toward around~$\tau=0$. 
This is due to an increasing level of shared large-scale topological features, e.g., LEV, toward the initial interaction, where the incoming gust vortex mainly interacts with the leading edge of the wing.
Past $\tau\approx 0$, we notice that $S_c$ with $\boldsymbol{u}_B={\boldsymbol{u}}-\boldsymbol{u_\infty}$ and $\boldsymbol{u}_B=\widetilde{\boldsymbol{u}}_{L}-\boldsymbol{u_\infty}$ begin to decrease.
This is caused by the differences in the evolution of the large-scale structures at different Reynolds numbers, as shown on the right of each subpanel in figure \ref{fig:fig7} that visualizes the Q criterion isosurface at $Re=600$ and 10,000 (for  $\sigma/c=0.05$) at $\tau=\tau_3$.
For instance, in the case with $G=2$~(figure~\ref{fig:fig7}$(a)$), the TEV is located further downstream at~$\tau=\tau_3$ for the $Re=$ 10,000 case compared to $Re=600$.
Nevertheless, the quantitative similarity remains high with $S_c$ above 0.6 even at~$\tau=2$.
The resemblance of the large-scale structures identified at higher Reynolds numbers to those observed at lower Reynolds numbers underscores the importance of laminar flow analysis for deepening our fundamental understanding of the dominant dynamics in extreme aerodynamic flows.
We note that the cosine similarity is a strict measure, since even the slightest difference in vortex position can significantly reduce the similarity.
Other approaches such as the optimal transport analysis~\citep{tran2025using} can also be considered for convective physics highlighting similarities in a more generalized manner.

\section{Concluding remarks}
\label{sec:conclusion}

This study revealed a high degree of similarity in extreme vortex-gust encounters by a square wing between $Re=600$ and 10,000.
We first presented similar coherent vortex cores initially formed via a significant vorticity production from the wing surface induced by the impacting gust vortex at both Reynolds numbers.
Through the use of scale decomposition, the large-scale flow structures were extracted at~$Re=$ 10,000. 
The identified large-scale vortical structures in the~$Re=$ 10,000 flow exhibit qualitative and quantitative similarity to those observed at~$Re=600$.
The similar pressure fields dominated by these large-scale vortical structures were also found, resulting in comparable lift fluctuations across the two Reynolds numbers.
Moreover, with the force element analysis, we presented that the prominent vortical structures responsible for the significant lift change during the extreme gust encounters are similar between the low and high Reynolds numbers.
Our findings also showed that the large-scale structures are the main contributors to significant lift peaks in the turbulent extreme flow. 
The current insights serve as a crucial bridge for understanding extreme gust encounters at high Reynolds numbers of at least~$\sim\mathcal{O}(10^4)$ through those with laminar cores, potentially paving the way to reduce the complexity of flow analysis, modeling, and control of massively separated flows in extreme aerodynamics. 
Future efforts at much higher Reynolds numbers are warranted to further examine the laminar core dynamics during violent extreme aerodynamic gust encounters.

\section*{Acknowledgments}
This work was supported by the US Department of Defense Vannevar Bush Faculty Fellowship (N00014-22-1-2798).
H.O. acknowledges partial support from Honjo International Scholarship Foundation.

\bibliographystyle{plainnat}
\bibliography{references}  

\end{document}